\documentclass[%
 reprint,
 amsmath,amssymb,
 aps,
]{revtex4-1}

\usepackage{graphicx}
\usepackage{dcolumn}
\usepackage{bm}
\usepackage[colorlinks, linkcolor=blue, anchorcolor=blue, citecolor=blue]{hyperref}
\usepackage{url}
\usepackage{amsmath,amssymb}
\usepackage{newtxmath}
\usepackage[mathlines]{lineno}
\usepackage{subfigure}
\usepackage{eurosym}

\begin{document}

\preprint{APS/123-QED}

\title{Spectrum of single-photon scattering in a strong-coupling hybrid optomechanical system}
\author{S. Y. Yang}
\affiliation{School of Physical Science and Technology, Southwest Jiaotong University, Chengdu 610031, China}
\author{W. Z. Jia}
\email{wenzjia@swjtu.edu.cn}
\affiliation{School of Physical Science and Technology, Southwest Jiaotong University, Chengdu 610031, China}
\author{Hao Yuan}
\affiliation{Key Laboratory of Opto-Electronic Information Acquisition and Manipulation of Ministry of Education, Anhui University, Hefei 230601, China}
\affiliation{CAS Key Laboratory of Quantum Information, University of Science and Technology of China, Hefei,
230026, China}
\date{\today}

\begin{abstract}
We analyze theoretically the single-photon excitation and transmission spectra of a strong-coupling hybrid optomechanics, where a two-level system (TLS) is coupled to the mechanical resonator (MR), generating the Jaynes-Cummings-type polariton doublets. In our model, both the optomichanical coupling and the TLS-MR  coupling are strong.  In this parameter region, the polaron-assisted excitation and reemission processes can strongly affect the single-photon excitation and output spectra of the cavity. We find that the fine structure around each sideband can be used to characterize the TLS-MR and the effective TLS-photon couplings, even at single-quantum level. Thus, the spectrum structures may make it possible to sensitively probe the quantum nature of a macroscopic mechanical element. We further provide a possible approach for tomographic reconstruction of the state of a TLS, utilizing the single-photon transmission spectra. 
\end{abstract}

\pacs{Valid PACS appear here}
\maketitle


\section{\label{Introduction}Introduction}

Optomechanical systems (OMSs) couple light and mechanical motion via radiation pressure. Cavity optomechanics has 
attracted extensive research interests for its potential applications in ultrasensitive measurements, quantum information 
processing, and implementation of novel quantum phenomena at macroscopic scales \cite{Kippenberg-Science2008,Marquardt-Physics2009,Barkai-Phys.Today2012,Aspelmeyer-Rev.Mod.Phys.2014}. Usually, the interaction 
between light and mechanical motion due to radiation pressure is intrinsically nonlinear but relatively weak (much 
smaller than the cavity linewidth). Thus in experiments to date, to obtain enhanced photon-phonon coupling strengths, 
strong optical drivings are usually used, but at the expense of making the effective interaction linear. Recently, extensive works have been focused on the single-photon strong-coupling regime, where the optomechanical coupling is comparable  to  the  cavity linewidth and  the  mechanical  frequency. Theoretical investigations have predicted some interesting 
phenomenon in this regime, including non-Gaussian steady states of the mechanical oscillator \cite{Nunnenkamp-PRL2011}, photon blockade \cite{Rabl-PRL2011,Liao-PRA2013,Xu-PRA2013}, photon-induced tunneling \cite{Xu-PRA2013}, scattering spectra with phonon sidebands \cite{Liao-PRA2012,Ren-PRA2013,Jia-PRA2013}, single-polariton physics in hybrid optomechanical sysyems \cite{Restrepo-PRL2014,Restrepo-PRA2017} and so on. 
Experimentally, this nonlinear regime has been reached in OMSs using ultracold atoms in optical resonators \cite{Murch-NatPhys2008,Brennecke-Science2008,Brooks-Nature2012}. Some other OMSs based on superconducting devices 
\cite{Teufel-Nature2011}, optomechanical crystal cavity \cite{Chan-Nature2011}, and spoke-anchored toroidal optical 
microcavity \cite{Verhagen-Nature2012} have also shown huge progress.  

It is known that mechanical resonators (MRs) can be used as powerful resources for quantum information and 
metrology, such as quantum memories and transducers connecting different quantum systems. There have been a great 
deal of experimental efforts to couple MRs to different kinds of two-level systems (TLSs), such as superconducting 
qubits \cite{Connell-Nature2010,Gustafsson-Science2014,Manenti-Nature2017,Moores-PRL2018,Chu-Science2017}, 
nitrogen-vacancy centers \cite{Arcizet-NatPhys2011,Kolkowitz-Science2012,Bennett-PRL2013}, and electron spin \cite{Rugar-Nature2004}. Typically, in quantum electromechanical devices with MRs strongly coupled to 
superconducting qubits \cite{Connell-Nature2010,Gustafsson-Science2014,Manenti-Nature2017,Moores-PRL2018,Chu-Science2017}, quantum control on phonons at single-phonon level can be realized. Thus in these strongly coupled TLS-MR systems, many phenomena in cavity QED system can be repeated utilizing atom-phonon interaction, and quantum information procession based on these structures become possible. 
A TLS can also be coupled to the mechanical element in an OMS, forming a kind of hybrid quantum system. It has been 
shown that this system can be used as transducers for long-distance quantum communication \cite{Stannigel-PRL2010}. Moreover, a TLS in an OMS can affect the ground state cooling of the MR \cite{Tian-PRB2011}, and induce 
single-phonon nonlinearities \cite{Ramos-PRL2013}. Two-color optomechanically induced transparency \cite{Wang-PRA2014} and tunable photon blockade effect \cite{Wang-PRA2015} in this system have also been studied. 
 
One can expect that if a hybrid OMS mentioned above works in the single-photon strong-coupling regime, even a driving field at single-photon level can be significantly affected by the state of the TLS-MR subsystem, leading to some observable features in the spectra of single-photon excitation and scattering. Thus, it is valuable to calculate analytically the single-photon spectra in this regime. However, to the best of our knowledge, this issue has not been studied before.   
In this paper, we analyze theoretically the single-photon excitation and transmission spectra of a hybrid OMS, where the 
MR is coupled to a TLS with Jaynes-Cummings-type interaction, utilizing a real-space approach \cite{Shen-OptLett2005,Shen-PRL2005}. Note that photon scattering problem in similar waveguide-emitter structures can also be dealt with some other methods  \cite{Zhou-PRL2008, Shi-PRB2009, Fan-PRA2010} and generalized to the case of nonlinear dispersion relation \cite{Zhou-PRL2008}.  In our hybrid system, both the optomechanical cavity  and the TLS-MR subsystem work in the strong coupling parameter region, where the optomechanical coupling is compared to the cavity linewidth and the mechanical frequency (so called single-photon strong coupling regime), and the TLS-MR coupling is lager than the decay of TLS  and the mechanical damping rate. We present the analytic 
solutions for the single-photon scattering problem. Our results show that the single-photon transmission spectra can be 
used to probe and characterize the TLS-MR interactions at single quantum level. It is known that probing the quantum 
nature of a macroscopic mechanical system has attracted a great deal of attention. The hybrid OMS investigated in this 
paper may provide a platform for this scope.
Another possible application of the single-photon transmission spectra is dispersive readout of the TLS as a qubit, 
because the qubit state can be mapped onto the cavity transmission peaks through strong nonlinear optomechanical 
interaction. 
 
The paper is organized as follows. In Sec.\ref {Hamiltonian-solutions}, we give a theoretical model, including the system 
Hamiltonian in Sec.\ref {Sys-Hamiltonian}, the energy-level structure of the TLS-MR subsystem in Sec.\ref {EL-structure}, and the scattering eigenstate in Sec.\ref {scattering-eigenstate}. The excitation spectra of the cavity are 
provided in Sec. \ref {Cavity excitation}. The single-photon transmission spectra are given in Sec. \ref {transmission-spectra}. In Sec. \ref {state-tomography}, we depict a method for dispersive readout of the TLS as a qubit. Finally, further 
discussions and conclusions are given in Sec. \ref {conclusion}.

\section{\label{Hamiltonian-solutions}Hamiltonian and solutions}

\subsection{\label{Sys-Hamiltonian}Hamiltonian of the system}

\begin{figure}[t]
    \centering
    \includegraphics[width=8cm]{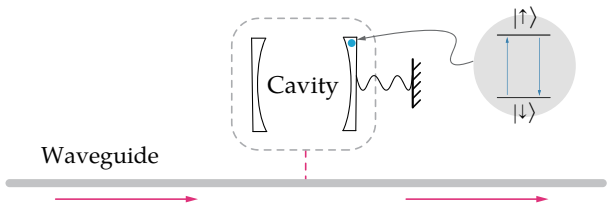}
    \caption{Schematic plot of the coupling system. A TLS is coupled to the moving mirror of an OMS. The optical cavity  is coupled to a waveguide, in which single photons propagate along the arrow direction.}
    \label{model}
\end{figure}

The system of interest consists of a hybrid OMS being side-coupled to a unidirectional single-mode waveguide, as 
shown schematically in Fig.~\ref{model}. The mechanical element of the OMS is coupled to a TLS, forming a cavity-QED-like subsystem. In our model, the waveguide-resonator system is assumed to be unidirectional,  such as a whispering-gallery-mode cavity coupled to a waveguide.  Also, the coupling configuration shown in Fig.~\ref{model} is equivalent to another case with the OMS being placed at one end of a semi-infinite waveguide, which is also widely adopted in experiments \cite{Aspelmeyer-Rev.Mod.Phys.2014}. The Hamiltonian of the system can be written as ($\hbar=1$)
\begin{equation}
\hat{H}=\hat{H}_{\mathrm{W}}+\hat{H}_{\mathrm{S}}+\hat{H}_{\mathrm{SW}},
\end{equation}
with
\begin{subequations}
    \begin{equation}
        \hat{H}_{\mathrm{W}}=
        \int \mathrm{d}x\,\hat{a}^{\dagger}(x)\left(-i\,v_{\mathrm{g}}\frac{\partial}{\partial x}\right)\hat{a}(x),
    \end{equation}
    \begin{eqnarray}
        \hat{H}_{\mathrm{S}}&=&
        \omega_{c}\hat{c}^{\dagger}\hat{c}+\omega_{b}\hat{b}^{\dagger}\hat{b}-g\hat{c}^{\dagger}\hat{c}\left(\hat{b}+
        \hat{b}^{\dagger}\right)
        \nonumber \\
        &&+\frac{\omega_{a}}{2}\hat{\sigma}_{z}+\lambda\left(\hat{b}\hat{\sigma}_{+}+\hat{b}^{\dagger}\hat{\sigma}_{-}\right),
        \label{systemHamiltonian}
    \end{eqnarray}
    \begin{equation}
        \hat{H}_{\mathrm{SW}}=
        V\int \mathrm{d}x\,\delta(x)\left[\hat{a}^{\dagger}(x)\hat{c}+\hat{a}(x)\hat{c}^{\dagger}\right].
    \end{equation}
\end{subequations}
$\hat{H}_{\mathrm{W}}$ denotes the waveguide optical modes, where $v_{\mathrm{g}}$ is the group velocity of the 
photons and $\hat{a}(x)$ [$\hat{a}^{\dagger}(x)$] is the bosonic operator annihilating (creating) a right-going photon at $x$. $\hat{H}_{\mathrm{S}}$ 
is the Hamiltonian of the isolated hybrid OMS with its mechanical element being coupled to a TLS. $\hat{c}$ ($\hat{c}^{\dagger}$)  and $\hat{b}$ ($\hat{b}^{\dagger}$) are the annihilation (creation) operators of the cavity and the 
mechanical modes, respectively. The Pauli operator $\hat{\sigma}_{z}=\left|\uparrow\right\rangle\left\langle\uparrow\right|-\left|\downarrow\right\rangle\left\langle\downarrow\right|$ is used to 
describe the TLS, while $\hat{\sigma}_{+}=\left|\uparrow\right\rangle\left\langle\downarrow\right|$ and $\hat{\sigma}_{-}=\left|\downarrow\right\rangle\left\langle\uparrow\right|$ are the ladder operators of the TLS. $\omega_{a}$ is the atomic 
transition frequency, $\omega_{c}$ is the cavity resonance frequency, and $\omega_{b}$ is the mechanical frequency. 
The interaction between the MR and the cavity field is described by the radiation-pressure interaction with single-photon coupling strength $g$. Here we assume the optomechanical coupling is compared to the cavity linewidth and the mechanical frequency i.e., the optomechanical subsystem works in the so called single-photon strong coupling regime. While the coupling between the MR and the TLS is described by the Jaynes-Cummings 
model with coupling strength $\lambda$. $\hat{H}_{\mathrm{SW}}$ describes the interaction between the waveguide 
and the optomechanical cavity, where $V$ is the corresponding coupling strength. The cavity-waveguide decay rate can be further 
defined as $\kappa=V^{2}/v_{\mathrm{g}}$. In experimental systems, the decay rates $\gamma_a$ of the TLS and $\gamma_b$ of the MR can be much smaller than the cavity-waveguide decay rate $\kappa$ \cite{Ramos-PRL2013,Chan-Nature2011,Lisenfeld-PRL2010}. Thus, during the time 
interval ${1}/{\kappa}\ll{t}\ll\min({1}/{\gamma_a},{1}/{\gamma_b})$, the input single photon has left the cavity, but the effects of 
the dissipation of the TLS-MR subsystem are still not obvious. Thus, in the analytic approach used to calculate single-photon scattering problem, we will ignore the damping processes of the TLS-MR subsystem.
  
\subsection{\label{EL-structure}Energy-level structure of the hybrid OMS}

Now we first calculate the energy levels of the hybrid OMS, which are useful for understanding the single-photon  scattering processes and constructing the scattering eigenstates. We introduce a unitary transformation $\hat{U}=\hat{U}_1\hat{U}_2$, with
\begin{subequations}
\begin{equation}
\hat{U}_1=\mathrm{exp}\left[\beta\hat{c}^{\dagger}\hat{c}\left(\hat{b}^{\dagger}-\hat{b}\right)\right],
\end{equation}
\begin{equation}
\hat{U}_2=\mathrm{exp}\left(-i\alpha\hat{c}^{\dagger}\hat{c}\hat{\sigma}_{y}\right),
\end{equation}
\end{subequations}
where  $\beta={g}/{\omega_b}$, $\alpha=\beta{\lambda}/{\omega_{a}}$. Physically, $\hat{U}_1$ is a photon-number-dependent displacement transformation of the MR states, and $\hat{U}_2$ generates a photon-number-dependent rotation in the Hilbert space of the TLS. By assuming  $\lambda\ll\omega_a,\omega_b$ (i. e. $\alpha\ll 1$), we can obtain
 \begin{equation}
 \hat{H}'_S=\hat{U}^{-1}\hat{H}_S\hat{U}\simeq \hat{H}'_{0}+\hat{H}'_{I},
 \label{HSys-prm}
 \end{equation}
with
\begin{subequations}
\begin{eqnarray}
    \hat{H}'_{0}&= &\omega_{c}\hat{c}^{\dagger}\hat{c}-\delta_1\left(\hat{c}^{\dagger}\hat{c}\right)^2+
    \frac{\omega_{a}}{2}\hat{\sigma}_{z}+\omega_{b}\hat{b}^{\dagger}\hat{b} \nonumber 
    \\
    &&+\lambda \left(\hat{b}\hat{\sigma}_{+}+\hat{b}^{\dagger}\hat{\sigma}_{-}\right), 
    \label{Hprime0}
\end{eqnarray}
\begin{equation}
    \hat{H}'_{I}=\delta_{2}\left(\hat{c}^{\dagger}\hat{c}\right)^2 \hat{\sigma}_{z}
    +\frac{\delta_2}{\beta}\hat{c}^{\dagger}\hat{c} \left(\hat{b}+\hat{b}^{\dagger}\right)\hat{\sigma}_{z}.
    \label{Hprime}
\end{equation}
\end{subequations}
Here $\delta_1={g^2}{/\omega_{b}}$ is the frequency shift for the cavity caused by the single-photon radiation pressure. $\delta_{2}=\alpha\beta\lambda={\beta}^2{\lambda}^2/{\omega_{a}}$ is an additional frequency shift for the cavity due to the effective interaction between a single photon and the TLS.  Alternatively,  the first term in Eq.~\eqref{Hprime} can be interpreted as the TLS transition being shifted by $2m^2\delta_{2}$ ($m=0,1,2,\cdots$ is the photon number). 

In $m$-photon subspace, the eigenstates of $\hat{H}'_{0}$  can be written as
\begin{equation}
\left|m\right>_c\left|{n\xi}\right>, 
\label{eigen_funH_0prim}
\end{equation}
where $\left|m\right>_c$ represents the number states of the cavity modes, and $\left|{n\xi}\right>$ is the eigenstates of the TLS-MR subsystem. Here, when the total excitation number of the TLS-MR subsystem $n\geqslant1$, we use $\xi=+,-$ to label the dressed-state pairs. And we use $n=0,\xi=\downarrow$ to label the ground state. Specifically,  the eigenstates $\left|{n\xi}\right>$ include the dressed eigenstates
\begin{subequations}
    \begin{equation}
       \left|n+\right>
        =\cos{\theta_n}\left|n-1\right>_b\left|\uparrow\right>+\sin{\theta_n}\left|n\right>_b\left|       
        \downarrow\right>, 
    \end{equation}
    \begin{equation}
       \left|n-\right>
        =-\sin{\theta_n}\left|n-1\right>_b\left|\uparrow\right>+\cos{\theta_n}\left|n\right>_b\left|
         \downarrow\right>, 
    \end{equation}
\end{subequations}
and the ground state $\left|0\downarrow\right>=\left|0\right>_b\left|\downarrow\right>$, where $\left|n\right>_b$ represent the number states of the mechanical modes. The mixing angle $\theta_n$ is defined as $\tan{2\theta_{n}}={2\lambda\sqrt{n}}/{\Delta}_{ab}$, and $\Delta_{ab}=\omega_a-\omega_b$ is the detuning between the TLS and the MR. The eigen energy of $\hat{H}'_{0}$ corresponding to eigenstate \eqref {eigen_funH_0prim}  can be written as
\begin{equation}
\epsilon_{m,n\xi}^{(0)}=m\omega_{c}-m^{2}\delta_{1}+\tilde{\epsilon}_{n\xi},
\end{equation}
where $\tilde{\epsilon}_{n\xi}$ is the eigen energy of the TLS-MR subsystem, and takes the form 
\begin{subequations}
\begin{equation}
\tilde{\epsilon}_{n\pm}=\left(n-\frac{1}{2}\right)\omega_{b}\pm\frac12\sqrt{\Delta^{2}_{ab}+4n\lambda^{2}},
\end{equation}
\begin{equation}
\tilde{\epsilon}_{0\downarrow}=-\frac{1}{2}\omega_{a}.
\end{equation}
\end{subequations}

Because $\alpha\ll 1$, we can consider $\hat{H}'_{I}$ as a perturbation part. According to first-order perturbation theory, the eigen energy of Hamlltonian \eqref {HSys-prm} can be written as
\begin{subequations}
\begin{equation}
    \epsilon_{m,n\pm}=
    \epsilon_{m,n\pm}^{(0)}\pm m^{2}\delta_2\cos2\theta_{n},
\end{equation}
\begin{equation}
    \epsilon_{m,0\downarrow}=\epsilon_{m,0\downarrow}^{(0)}-m^{2}\delta_2.
\end{equation}
\end{subequations}
 The corresponding eigenequation is
\begin{equation}
  \hat{H}'_{S} \left|m\right>_c\left|{n\xi}\right> =\epsilon_{m,n\xi} \left|m\right>_c\left|{n\xi}\right>.
\end{equation}
Using the relation $\hat{H}'_S=\hat{U}^{-1}\hat{H}_S\hat{U}$, we can write down the following eigenequation 
\begin{equation}
    \hat{H}_{S} \left|m\right>_c\left|\widetilde{n\xi}(m)\right>
    =\epsilon_{m,n\xi} \left|m\right>_c\left|\widetilde{n\xi}(m)\right>.
\end{equation}
Here $|\widetilde{n\xi}(m)\rangle=\hat{U}(m)\left|{n\xi}\right>$, with $\hat{U}(m)=\left<m\right|\hat{U}\left|m\right>_c$. Note that the rotation parameter $\alpha\ll 1$, thus $|\widetilde{n\xi}(m)\rangle$ can be approximately looked on as an $m$-photon displaced dressed state of the TLS-MR subsystem.

\begin{figure*}[t]
    \centering
    \includegraphics[width=12cm]{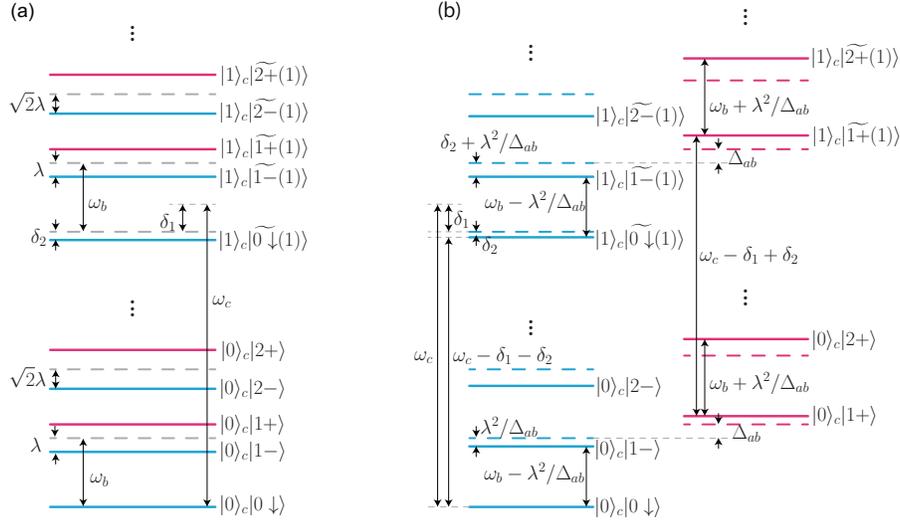}
    \caption{The energy-level structure of the hybrid OMS in the single-photon strong-coupling regime. (a) The spectrum  for
the case of zero detuning $\Delta_{ab}=0$ between the TLS and MR. (b) The spectrum  for
the case of large detuning  $\Delta_{ab}\gg\lambda$.}
    \label{ELStructure}
\end{figure*}

Here we are especially interested in the cases of zero TLS-MR detuning $\Delta_{ab}=0$ and the dispersive regime 
$|\Delta_{ab}|\gg\lambda$.  When  $\Delta_{ab}=0$, the eigenstates of the  TLS-MR subsystem exhibit the polariton doublets
\begin{equation}
\left|{n\pm}\right>
=\frac{1}{\sqrt{2}} \left(\left|n\right>_b\left|       
\downarrow\right>\pm\left|n-1\right>_b\left|\uparrow\right>\right).
\label{dressed-stateR}
\end{equation}   
The corresponding eigen energies
\begin{equation}
\tilde{\epsilon}_{n\pm}=\left(n-\frac{1}{2}\right)\omega_{b}\pm\sqrt{n}\lambda,
\end{equation}
give rise to the so-called Jaynes-Cummings ladder.
For large TLS-MR detuning, $|\Delta_{ab}|\gg\lambda$, the eigenstates of the $n$-excitation manifolds (with small $n$) take the form
\begin{subequations}
\begin{align}
\left|n+\right\rangle\simeq\left\vert n-1\right\rangle_b\left\vert\uparrow\right\rangle+\sqrt{n}\frac {\lambda}{\Delta_{ab}}\left\vert n\right\rangle_b\left\vert\downarrow\right\rangle,
\end{align}
\begin{align}
\left\vert n-\right\rangle\simeq\left|n\right\rangle_b\left\vert\downarrow\right\rangle -\sqrt{n}\frac{\lambda}{\Delta_{ab}}\left|
n-1\right\rangle_b\left\vert\uparrow\right\rangle,
\end{align}
\label{dressed-stateLD}
\end{subequations}
with the corresponding eigenenergies
\begin{subequations}
\begin{equation}
\tilde{\epsilon}_{n+}\simeq\frac{1}{2}\omega_{a}+\frac{\lambda^2}{\Delta_{ab}}+\left(n-1\right)\left(\omega_b+\frac{\lambda^2}{\Delta_{ab}}\right),
\end{equation}
\begin{equation}
\tilde{\epsilon}_{n-}\simeq-\frac{1}{2}\omega_{a}+n\left(\omega_b-\frac{\lambda^2}{\Delta_{ab}}\right).
\end{equation}
\end{subequations}
The energy-level structure of the hybrid OMS is plotted in Fig.~\ref{ELStructure}. 

\subsection{\label{scattering-eigenstate}Scattering eigenstates}
We assume that initially the TLS-MR subsystem is prepared in its eigenstate $\left|n_{0}\xi_{0}\right>$, the cavity is empty, and a single photon with frequency $\omega_k=v_{\mathrm{g}}k$ comes from the left, where $k$ is the wave vector of the photon. In this case, the total energy of the system is $E=\omega_k+\epsilon_{0,n_{0}\xi_{0}}$. Because the total photon number $\hat{N}=\int \mathrm{d}x\hat{a}^{\dagger}(x)\hat{a}(x)+\hat{c}^{\dagger}\hat{c}$ is a conserved quantity,  we assume that the stationary state $\left|E\right>$ of the system can be expanded in the single-photon subspace as 
\begin{align}
    \left|E\right>=
    &\sum_{n\xi}\int\mathrm{d}x\varphi_{n\xi,n_0\xi_0}(x)\hat{a}^{\dagger}(x)\left|\emptyset\right>\left|0\right>_{c}\left|n\xi\right>
    \notag \\
    &+\sum_{n\xi}e_{n\xi,n_0\xi_0}\hat{c}^{\dagger}\left|\emptyset\right>\left|0\right>_{c}\left|\widetilde{n\xi}\left(1\right)\right>,
    \label{Int-eigenstate}
\end{align}
where $\left|\emptyset\right>\left|0\right>_{c}$ is the vacuum state, with zero photon in both the waveguide and the cavity. $\varphi_{n\xi,n_0\xi_0}(x)$ represents the single-photon wave function of the waveguide mode. $e_{n\xi,n_0\xi_0}$ is the single-photon excitation amplitude of the cavity.

The eigenequation 
\begin{equation}
\hat{H}\left|E\right>=E\left|E\right>
\end{equation}
satisfied by the stationary state $\left|E\right>$ gives the following set of equations of motion 
\begin{subequations}
    \begin{eqnarray}
            -i\,v_{\mathrm{g}}\frac{\partial\varphi_{n\xi,n_0\xi_0}(x)}{\partial x}
        +\delta(x)V\sum_{n'\xi'}e_{n'\xi',n_0\xi_0}\langle n\xi|\widetilde{n'\xi'}
        (1)\rangle
        \nonumber 
        \\
       =\left(E-\epsilon_{0,n\xi}\right)\varphi_{n\xi,n_0\xi_0}(x),
        \nonumber 
        \\
        \label{MotionEq1}
      \end{eqnarray}
      \begin{eqnarray}
           \sum_{n'\xi'}\left(E-\epsilon_{1,n'\xi'}\right)e_{n'\xi',n_0\xi_0}\langle n\xi|
            \widetilde{n'\xi'}(1)\rangle
            \nonumber
            \\
           =V\int\mathrm{d}x\delta(x)\varphi_{n\xi,n_0\xi_0}(x).
       \label{MotionEq2}
     \end{eqnarray}
\end{subequations}

The single-photon wave function $\varphi_{n\xi,n_0\xi_0}(x)$ should take the form
\begin{eqnarray}
    \varphi_{n\xi,n_0\xi_0}(x)&=&\theta(-x)\delta_{n\xi,n_0\xi_0}e^{ikx}
    \nonumber 
    \\
    &&+\theta(x)t_{n\xi,n_0\xi_0}e^{i\,\left(k-\frac{\epsilon_{0,n\xi}-\epsilon_{0,n_{0}\xi_{0}}}{v_{\mathrm{g}}}\right)x},
    \label{WaveFunction}
\end{eqnarray}
where  $t_{n\xi,n_0\xi_0}$ is the transmission amplitude, and $\theta(x)$ denotes the Heaviside step function. 
We point out that the ansatz \eqref{WaveFunction} is valid if the condition $\gamma_a,\gamma_b\ll\kappa$ is satisfied.  Thus during the time duration ${1}/{\kappa}\ll{t}\ll\min({1}/{\gamma_a},{1}/{\gamma_b})$ we are interested in, the single photon leaks completely out of the cavity and meanwhile the decays of the TLS and MR are negligible.
\begin{figure*}[t]
\centering
\includegraphics[width=16cm]{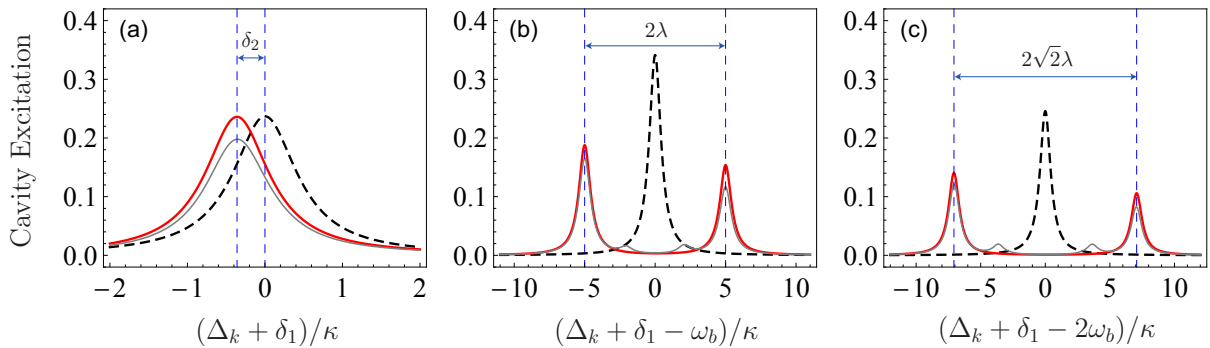}
\caption{Fine structures of cavity excitation spectrum around different sidebands as a function of detuning $\Delta_k$.  The parameters are $\Delta_{ab}=0$, $\kappa=0.01\omega_b$, and $g=1.2\omega_b$ for all curves. The thick curves are plotted with the analytical result in Eq.~\eqref {cavity-excitation}.  The thick solid lines show the cavity excitation spectrum for $\lambda=5\kappa$, and the thick dashed lines for $\lambda = 0$. The thin gray solid lines are numerical results for the $\lambda=5\kappa$ case, where the decay rates of the TLS-MR system are included and a non-zero temperature is assumed, with $\gamma_a=10^{-4}\omega_b$, $\gamma_b=10^{-5}\omega_b$, and $n_a=n_b=0.1$.}
\label{excitationR}
\end{figure*}

Substituting Eq.~\eqref{WaveFunction} into Eqs.~\eqref{MotionEq1} and \eqref{MotionEq2}, we can obtain
\begin{subequations}
    \begin{equation}
        e_{n\xi,n_0\xi_0}(\omega_k)=\frac{V\langle\widetilde{n\xi}(1)|n_{0}\xi_{0}\rangle}{\widetilde{\Delta}_{n\xi,n_0\xi_0}
        (\omega_k)},\label{ExcitationAmp}
    \end{equation}
    \begin{equation}
        t_{n\xi,n_0\xi_0} (\omega_k) =\delta_{n\xi,n_0\xi_0}-i\kappa\sum_{n'\xi'}\frac{\langle n\xi|
        \widetilde{n'\xi'}(1)\rangle\langle\widetilde{n'\xi'}(1)|n_{0}\xi_{0}\rangle}{\widetilde{\Delta}_{n'\xi',n_{0}\xi_{0}}
        (\omega_k)},
        \label{TransmissionAmp}
    \end{equation}
\end{subequations}
with
\begin{equation}
    \widetilde{\Delta}_{n\xi,n_0\xi_0}(\omega_k)=\omega_k+\epsilon_{0,n_0\xi_0}-\epsilon_{1,n\xi}+i\frac{\kappa}{2}.
    \label{EffectiveDetuning}
\end{equation}
The explicit formulas for the overlaps like $\langle n\xi|\widetilde{n'\xi'}(1)\rangle$ are calculated in Appendix \ref{Uele-Cal}. 

The physical picture of single-photon scattering process described by the scattering eigenstates can be summarized from Eqs.~\eqref{WaveFunction}-\eqref{EffectiveDetuning}. After entering the cavity, the incident photon can excite the hybrid system from the initial state $|0\rangle_c|n_0\xi_{0}\rangle$ to the upper state 
$|1\rangle_c|\widetilde{n'\xi'}(1)\rangle$.  The resonance condition of this process is $\omega_k=\epsilon_{1,n'\xi'}-\epsilon_{0,n_0\xi_0}$. Then the upper state finally decays into the state $|0\rangle_c|n\xi\rangle$, resulting in a re-emitted photon with frequency $\omega_k-({\epsilon_{0,n\xi}-\epsilon_{0,n_{0}\xi_{0}}})$. Clearly, when the initial and the final states of the system are the same, the photon is scattered elastically. In this case, the photon transmission is the consequence of interference between the direct transmission and the cavity reemission, as shown by Eq.~\eqref{TransmissionAmp}, When the initial and the final states are different, the inelastic scattering happens, and the transmission is simply the cavity reemission. 

\section{\label{Cavity excitation}Cavity excitation}

If the TLS-MR subsystem is initially prepared in its ground state $\left|0\downarrow\right>$, the corresponding cavity excitation can be defined as
\begin{equation}
\sum_{n\xi}\left|e_{n\xi,0\downarrow}(\omega_k)\right|^2,
\label{cavity-excitation}
\end{equation}
which is in proportion to the probability for the cavity being excited by an input single photon with frequency $\omega_k$, when the interaction between the photon and the system happens.  We plot the cavity excitation spectrum as the functions of the photon-cavity detuning $\Delta_k=\omega_k-\omega_c$ when the system entering the single-photon strong-coupling regime, as shown in Fig. \ref{excitationR}. In the resolve sideband limit $\kappa\ll\omega_b$, the cavity response shows several resolved resonances, corresponding to different excitation pathways from the ground state $|0\rangle_c|{0\downarrow}\rangle$ to the one-photon state $|1\rangle_c|\widetilde{n\xi}(1)\rangle$. They are resonant if the frequency of the  single photon $\omega_k$ matches 
\begin{equation}
\omega_k=\epsilon_{1,n\xi}-\epsilon_{0,0\downarrow}.
\label{Excit-match-condition}
\end{equation}
The resonances are weighted by a factor $|\langle\widetilde{n\xi}(1)\left|{0\downarrow}\right\rangle|^2$ and the widths are $\kappa$. 

When the TLS is
in resonance with the MR, i.e., $\Delta_{ab}= 0$, the condition \eqref{Excit-match-condition} can be further written as
\begin{equation}
\Delta_k=\begin{cases}-\delta_1-\delta_2 & n=0, 
\\
 -\delta_1+n\omega_b\pm\sqrt{n}\lambda & n>0.
\end{cases}
\end{equation}
Specifically, the peak located at $\Delta_k=-\delta_1-\delta_2$ corresponds to the transition pathway 
$|0\rangle_c|{0\downarrow}\rangle\rightarrow|1\rangle_c|\widetilde{0\downarrow}(1)\rangle$ [shown by the thick solid line in 
Fig.~\ref{excitationR}(a)], and the double peaked structures centered at $\Delta_k=-\delta_1+n\omega_b$ with splitting width $2\sqrt{n}\lambda$ correspond to the transition pathways  $|0\rangle_c|{0\downarrow}\rangle\rightarrow|1\rangle_c|\widetilde{n\pm}(1)\rangle$ [shown by the solid lines in Figs.~\ref{excitationR}(b) and \ref{excitationR}(c)].
Clearly, these double peaks characterize the polariton doublets [see Eq.~\eqref{dressed-stateR}] of the TLS-MR 
subsystem.  Moreover, at each sideband, the double peaks exhibit asymmetric spectral structure.  Note that when a cavity QED system  works in the ultra-strong coupling regime, asymmetry of the Rabi splitting spectrum will appear due to the effects of anti-rotating terms \cite{Cao-NewJournalofPhys2011}.  But in our case, we have assume that the atom-MR coupling rate is much less than the atomic and the mechanical frequencies. thus the effects of anti-rotating terms can be omitted. The asymmetric spectral structure in our case is resulted from the strong nonlinear interaction between photons and phonons.  The thick dashed lines in Figs.~\ref{excitationR}(a)-\ref{excitationR}(c) show the cavity excitation when the 
TLS is not included (i.e., $\lambda=0$) for comparison. One can find that in the $\lambda=0$ case the sideband peaks 
appear at  frequency $-\delta_1+n\omega_b$ ($n=0,1,2\cdots$), which has been well investigated previously 
\cite{Nunnenkamp-PRL2011,Rabl-PRL2011,Liao-PRA2012,Ren-PRA2013}. Compared with $\lambda=0$ case, the first (labeled from left to right) resonance peak undergoes a red shift $\delta_2$ [see the thick dashed and solid lines in 
Fig.~\ref{excitationR}(a)], which can be resolved from the Lorentzian spectrum when $
\delta_2 \sim\kappa$, i.e., the condition $\lambda\sim{\sqrt{\omega_a\kappa}/\beta}$ is satisfied. As can be seen from the Hamiltonian \eqref {Hprime}, this shift characterizes the strength of the effective photon-TLS interaction mediated by the MR.   

\begin{figure*}[t]
\centering
\includegraphics[width=16cm]{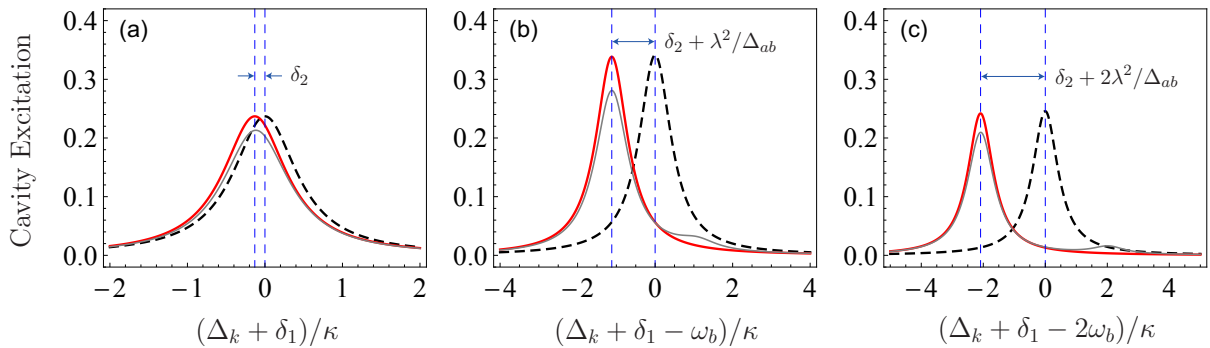}
\caption {
Fine structure of cavity excitation spectrum around different sidebands as a function of detuning $\Delta_k$.     
$\Delta_{ab}=0.1\omega_b$, $\kappa=0.001\omega_b$, and $g=1.2\omega_b$ for all curves.  The thick curves are plotted with the analytical result in Eq.~\eqref {cavity-excitation}.  The thick solid lines show the cavity excitation spectrum for $\lambda=0.1\Delta_{ab}$, and the thick dashed lines for $\lambda = 0$. The thin gray solid lines are numerical results for the $\lambda=0.1\Delta_{ab}$ case, where the decay rates of the TLS-MR system are included and non-zero temperature is assumed, with $\gamma_a=0.01\lambda$, $\gamma_b=10^{-5}\omega_b$, and $n_a=0.077, n_b=0.1$.}
\label{excitationLD}
\end{figure*}

The cavity excitation for single-photon scattering defined in Eq.~\eqref{cavity-excitation} is equivalent to the steady-state 
mean photon number $\langle\hat{c}^{\dagger}\hat{c}\rangle_{\mathrm{ss}}$ at zero temperature when the cavity 
is driven by a weak continuous wave laser. Thus, to verify our analytical calculations given above, we numerically 
simulate the steady-state mean photon number of the driven cavity utilizing the master equation method. In our simulation, the decay rates $\gamma_a$ and $\gamma_b$ are also included. According to Refs.~\cite{Ramos-PRL2013,Chan-Nature2011,Lisenfeld-PRL2010}, we let $\gamma_a=10^{-4}\omega_b,\gamma_b=10^{-5}\omega_b$. 
We find that for the zero-temperature case with $n_{a} =n_{b} = 0$ (Here the Bose occupation numbers are defined as $n_{a,b}^{-1}=e^{\hbar\omega_{a,b}/k_{B}T}-1$ ), the numerical results 
are in good agreement with the analytical ones (The numerical results for this case are not shown in the figures 
because the numerical curves are almost coincided with the theoretical ones). We also simulate the excitation spectra 
for $n_{a},n_{b}>0$ case, as shown by the thin gray solid lines in Figs.~\ref{excitationR}(a)-\ref{excitationR}(c). For this nonzero-temperature case, 
the excited states of the Jaynes-Cummings ladder will be populated, which can provide more possible transition 
pathways \cite{Fink-PRL2010}. As a result, additional peaks will appear in the spectra. For example, the new peaks 
appearing in the thin gray solid lines in Fig.~\ref{excitationR}(b) and \ref{excitationR}(c) are related with the transitions 
$|0\rangle_c|{{1\xi}}\rangle\rightarrow|1\rangle_c|\widetilde{2\xi}(1)\rangle$ and  $|0\rangle_c|{{1\xi}}\rangle\rightarrow|1\rangle_c|\widetilde{3\xi}(1)\rangle$, respectively.  
Meanwhile, compared with zero-temperature case, the already existing peaks corresponding 
to  $|0\rangle_c|{0\downarrow}\rangle\rightarrow|1\rangle_c|\widetilde{n\xi}(1)\rangle$ transitions will decrease, as shown 
in Figs.~\ref{excitationR}(a)-\ref{excitationR}(c),
because the population of the ground state  $|0\rangle_c|{0\downarrow}\rangle$ becomes reduced at nonzero temperatures.   

Now we consider the large TLS-MR detuning case. Under condition  $\omega_{i}\gg|\Delta_{ab}|\gg\lambda$ ($i=a,b$), we have
\begin{equation}
\left|\frac{\langle\widetilde{n+}(1)|0\downarrow\rangle}{\langle\widetilde{n-}(1)|0\downarrow\rangle}\right|^2\simeq{n}\frac{{\lambda}^2}{{\Delta}_{ab}^2},
\end{equation}
which is much smaller than one when the excitation number $n$ is small. Thus, in the large detuning case, the processes $|0\rangle_c|{0\downarrow}\rangle\rightarrow|1\rangle_c|\widetilde{n-}(1)\rangle$ are dominant,  whereas the  transition pathways $|0\rangle_c|{0\downarrow}\rangle\rightarrow|1\rangle_c|\widetilde{n+}(1)\rangle$ are highly suppressed. From Eq.~\eqref {Excit-match-condition}, the resonance peaks characterizing the transitions $|0\rangle_c|{0\downarrow}\rangle\rightarrow|1\rangle_c|\widetilde{n-}(1)\rangle$ appear at $\omega_k=\epsilon_{1,n-}-\epsilon_{0,0\downarrow}$, which can be further written as
\begin{equation}
\Delta_k\simeq -\delta_1-\delta_2 +n\left(\omega_b-\frac{{\lambda}^2}{\Delta_{ab}}\right).
\end{equation}
We can see that when a TLS couples to the MR with large detuning, the resonance  
peaks are spaced by the dispersively shifted frequency $\omega_b-{{\lambda}^2}/{\Delta_{ab}}$ of the phonon-like 
polariton. Namely, the spectra characterize the polariton-assisted cavity excitation. Also, the additional frequency shift 
$\delta_2$ results from the effective photon-TLS interaction. Above analysis can be verified by the thick solid lines in Figs.~\ref{excitationLD}(a)-\ref{excitationLD}(c). The excitation 
spectra for the TLS free case are also plotted (thick dashed lines), where the resonance peaks appear at 
$\Delta_k= -\delta_1 +n\omega_b$. 

The thin gray solid lines in Figs.~\ref{excitationLD}(a)-\ref{excitationLD}(c) are numerical results for the large TLS-MR detuning case, where the decay rates of the TLS-MR system are included and non-zero temperature is assumed. We find that similar to the resonance case, at nonzero temperatures, the peaks corresponding to  $|0\rangle_c|{0\downarrow}\rangle\rightarrow|1\rangle_c|\widetilde{n-}(1)\rangle$ transition decrease and additional peaks corresponding to  $|0\rangle_c|{1+}\rangle\rightarrow|1\rangle_c|\widetilde{n+}(1)\rangle$ transition can be observed [see the peaks at $\Delta_k=-\delta_1+\delta_2+n(\omega_b+\lambda^2/\Delta_{ab})$ in the thin gray lines in Figs.~\ref{excitationLD}(b) and \ref{excitationLD}(c)], because the population of the ground state  $|0\rangle_c|{0\downarrow}\rangle$ becomes reduced and higher dressed states get populated. 

\section{\label{transmission-spectra}Single-photon transmission spectra}

\begin{figure*}[t]
\centering
\includegraphics[width=16cm]{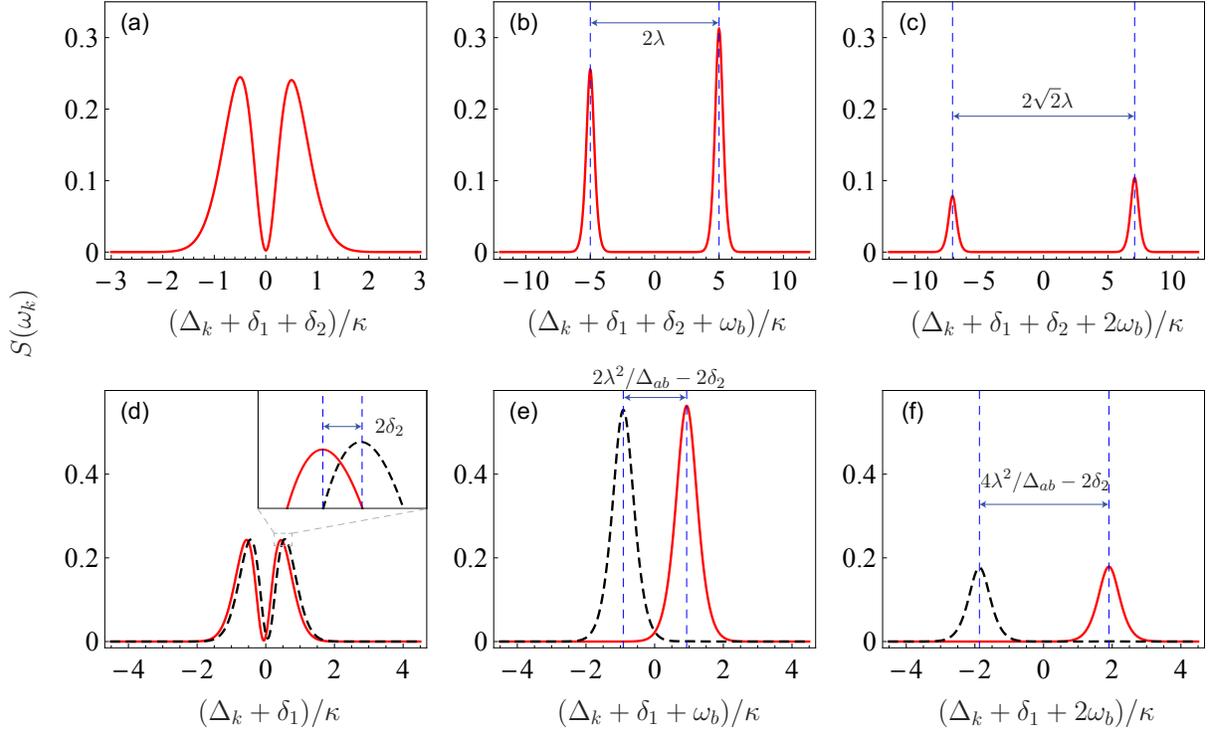}
\caption {Single-photon transmission spectrum around different sidebands as a function of detuning $\Delta_k$.  $g=0.8\omega_b$, and $d=\kappa$ for all curves. (a)-(c): The TLS-MR resonance case with $\Delta_{ab}=0$, $\kappa=0.01\omega_b$, and $\lambda=5\kappa$; (d)-(f): The large TLS-MR detuning case with $\Delta_{ab}=0.1\omega_b$, $\kappa=0.001\omega_b$ and  $\lambda=0.1\Delta_{ab}$. The solid lines in (d)-(f) correspond to the initial state $\left|\downarrow\right\rangle$ of the TLS, and the dashed lines correspond to the state $\left|\uparrow\right\rangle$. }
\label{transmission-spectrum}
\end{figure*}

Using the scattering eigenstates given in Sec.~\ref{scattering-eigenstate}, we can construct the scattering matrix and further calculate the transmission spectrum of a single-photon. We consider an incident photon with an Gaussian-type spectral amplitude $f(\omega_k)=(2/\pi d^2)^{1/4}\mathrm{exp}\left[-(\omega_k-\omega_0)^2/d^2\right]$, where $\omega_0$ is the center frequency, and $d$ is spectrum width of the photon. We assume the TLS-MR subsystem is initially in a superposition state $\sum_{n_0\xi_0}C_{n_0\xi_0}|n_0\xi_0\rangle$. The corresponding spectrum of the transmitted photon is
\begin{widetext}
    \begin{equation}
        S\left(\omega_{k}\right)=\sum_{n\xi}\Biggl|\sum_{n_0\xi_0}C_{n_0\xi_0}f \left(\omega_{k}+\epsilon_{0,n\xi}-
        \epsilon_{0,n_{0}\xi_{0}}\right)t_{n\xi,n_0\xi_0} \left(\omega_{k}+\epsilon_{0,n\xi}-\epsilon_{0,n_{0}\xi_{0}}\right)
        \Biggr|^2, 
        \label{ScatteringSpectrum}    
    \end{equation}
\end{widetext}
(see Appendix \ref {Cal-OutputSpectrum} for details), which represents the probability density for finding a single photon with frequency $\omega_{k}$ in the transmission field.

We first consider the case that the TLS-MR 
subsystem satisfies the exact resonance condition $\Delta_{ab}=0$. We assume the TLS-MR subsystem is initially 
prepared in its ground state $|0\downarrow\rangle$, and a single-photon wave packet with center 
frequency $\omega_0=\omega_c-\delta_1-\delta_2$ incidents. The width $d$ of the wave packet is chosen to be much 
less than the mechanical frequency $\omega_b$.  Thus, the single photon will first excite resonantly the transition $|0\rangle_c|{0\downarrow}\rangle\rightarrow|1\rangle_c|\widetilde{0\downarrow}(1)\rangle$. Subsequently, the photon in 
the cavity decay into the waveguide through transitions $|1\rangle_c| {\widetilde{0\downarrow}(1)}\rangle\rightarrow|0\rangle_c|n\xi\rangle$. 
Specifically, the peak at $\Delta_k=-\delta_1-\delta_2$, which results from the elastic scattering 
related to the transition pathway $|0\rangle_c|{0\downarrow}\rangle\rightarrow|1\rangle_c|\widetilde{0\downarrow}(1)\rangle\rightarrow|0\rangle_c|{0\downarrow}\rangle$, is split due to the destructive interference between direct 
transmission and cavity photon reemission, as shown in Fig.~\ref{transmission-spectrum}(a). And the inelastic scatted  
photons can generate peaks at frequency $\Delta_k=-\delta_1-\delta_2-n\omega_b\mp\sqrt{n}\lambda$ ($n=1,2\cdots$) 
in the output spectrum, as shown in Figs.~\ref{transmission-spectrum}(b)-\ref{transmission-spectrum}(c). 
 
Now we discuss the transmission spectra when the TLS is coupled to the MR with large detuning 
$|\Delta_{ab}|\gg\lambda$. If the TLS-MR subsystem is initially prepared in its ground state 
$\left|0\downarrow\right\rangle$,  an incident wave packet with center frequency 
$\omega_0=\omega_c-\delta_1-\delta_2$ mainly generates the scattering processes related to the transition pathways 
$|0\rangle_c|{0\downarrow}\rangle\rightarrow|1\rangle_c|\widetilde{0\downarrow}(1)\rangle\rightarrow|0\rangle_c|{n-}\rangle$, 
where the state of the TLS is almost unchanged after scattering.  While the processes  
$|0\rangle_c|{0\downarrow}\rangle\rightarrow|1\rangle_c|\widetilde{0\downarrow}(1)\rangle\rightarrow|0\rangle_c|{n+}\rangle$, 
corresponding approximatively to a qubit flip, is highly suppressed.  Correspondingly, we may find resonance peaks at 
$\Delta_k=-\delta_1-\delta_2-n(\omega_b-\lambda^2/\Delta_{ab})$, as shown by the solid lines in Figs. \ref{transmission-spectrum}(d)-\ref{transmission-spectrum}(f). 
Note that the peak at $\Delta_k=-\delta_1-\delta_2$ charactering elastic scattering (i.e., $n=0$ case) is split due to destructive interference 
[solid line in Fig. \ref{transmission-spectrum}(d)]. If the TLS-MR subsystem is initially prepared in the state 
$|0\uparrow\rangle$,  for an incident wave packet with center frequency $\omega_0=\omega_c-\delta_1+\delta_2$, the 
scattering processes related to the transition pathways $|0\rangle_c|{1+}\rangle\rightarrow|1\rangle_c|\widetilde{1+}(1)\rangle\rightarrow|0\rangle_c|{n+}\rangle$ is dominant, resulting in the resonance peaks appearing at 
$\Delta_k=-\delta_1+\delta_2-n(\omega_b+\lambda^2/\Delta_{ab})$, as shown by the dashed lines in 
Figs. \ref{transmission-spectrum}(d)-\ref{transmission-spectrum}(f). 

In summary, under the large detuning condition, the TLS-MR subsystem can be looked on as an  
effective MR, with disperse-shifted frequencies $\omega_b\pm\lambda^2/\Delta_{ab}$, depending on the TLS being in 
its ground or excited states. Compared with the standard OMS (i.e., $\lambda=0, \delta_2=0$ case with peaks appearing at $\Delta_k=-\delta_1-n\omega_b$), the position of the $n$-th sideband in the transmission 
spectra will shifted by $\pm({n}\lambda^2/\Delta_{ab}-\delta_2)$, depending on the state of the TLS. Thus, similar to 
cavity QED or circuit QED system \cite{Alexandre-PRA2004, Gu-PhysRep2017, Blais-arXiv2020}, the single-photon transmission spectrum in the TLS-MR 
dispersive regime is advantageous for readout of the TLS as a qubit.  And we will discuss this issue in detail in the next 
section.   

\begin{figure*}[t]
\centering
\includegraphics[width=10cm]{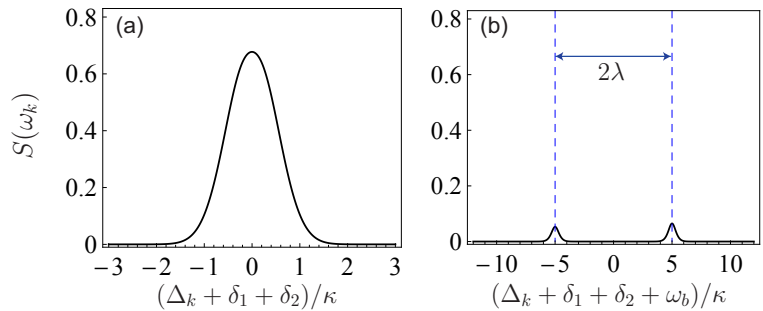}
\caption {Single-photon transmission spectrum around different sidebands as a function of detuning $\Delta_k$. The optomechanical coupling strength is $g=0.2\omega_b$, other parameters are the same as those used in Figs.~\ref{transmission-spectrum}(a)-\ref{transmission-spectrum}(c).}
\label{weakcoupling}
\end{figure*}

Finally, we emphasize that to obtain remarkable spectral peaks of phonon sidebands like those discussed above, the single-photon strong-coupling condition 
$g\sim\omega_b$  is required.  In Fig.~\ref{weakcoupling}, we plot the transmission spectrum for a relatively small optomechanical coupling strength for 
comparison. We take the case of on-resonance TLS-MR interaction as an example, the optomechanical coupling strength is chosen as $g=0.2\omega_b$, 
other parameters are the same as those used in Figs.~\ref{transmission-spectrum}(a)-\ref{transmission-spectrum}(c). One can see that the main peak at $\Delta_k=-\delta_1-\delta_2$ is 
dominant [Fig.~\ref{weakcoupling}(a)], while the double peaks at the first sideband become very weak [Fig.~\ref{weakcoupling}(b)]. Note that the second and 
higher sidebands almost vanish under this coupling strength, thus they are not shown in the figure. In fact, when the optomechanical coupling strength is 
smaller than this value, the transmission spectra become almost the same as those of a usual cavity with fixed mirrors. 

\section{\label{state-tomography}Quantum state tomography via the cavity output spectra}

\begin{figure*}[t]
\centering
\includegraphics[width=16cm]{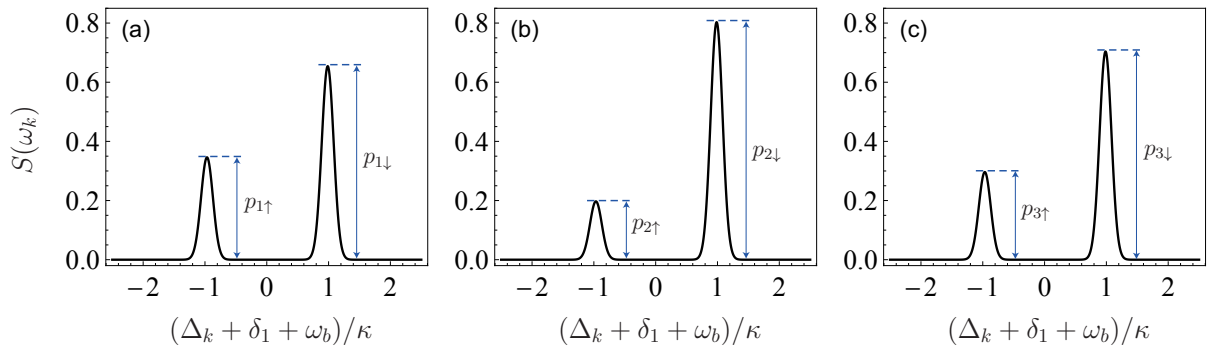}
\caption {Single-photon transmission spectrum around the first sideband as a function of detuning $\Delta_k$. The width of the single-photon wave packet is $d=0.2\kappa$. Other parameters are the same as in Figs.~\ref{transmission-spectrum}(d)-\ref{transmission-spectrum}(f). Panels (a)-(c) correspond to the detected states $\hat{\rho}_1$, $\hat{\rho}_2$, and $\hat{\rho}_3$, respectively.}
\label{state-readout}
\end{figure*}

Here, as an example, we propose a feasible proposal for quantum state tomography (QST) based on measurements in the mutually unbiased bases (MUBs) \cite{Wootters-AnnPhys1989, Ivonovic-JPhysA1981, Han-QuantumInfProcess2015, Klimov-PRA2013,Yan-PRA2010, Klimov-PRA2008, Adamson-PRL2010, Rohling-PRB2013, Yuan-NewJPhys2016}, utilizing our hybrid OMS as a physical implementation. It has been proved that the measurements in the MUBs provide a minimal and optimal way to realize QST (called MUBs-QST hereafter) in the sense of maximizing information extraction from each measurement and minimizing the effects of statistical errors in the measurements \cite{Wootters-AnnPhys1989}.

We first provide a brief review of MUBs-QST for a TLS. In terms of the MUBs, the density operator describing an arbitrary quantum state of a TLS can be represented as \cite{Ivonovic-JPhysA1981}
\begin{equation}
\hat{\rho}=\sum_{s=1}^{3}\sum_{l=\downarrow,\uparrow}p_{sl}\hat{P}_{sl}-\hat{I},
\label{density-matrix-in-MUBs}
\end{equation}
where the MUBs-projector $\hat{P}_{sl}=|\psi_{sl}\rangle\langle \psi_{sl}|$ defines a complete set of projective 
measurements, $p_{sl}$ is the probability of projecting $\hat{\rho}$ onto the basis state $|\psi_{sl}\rangle$ of the $s$-th 
MUB, and $\hat{I}$ is an identity operator. The $s$-th MUB $|\psi_{sl}\rangle$ can be transformed from the standard 
computational basis $\left\vert l\right\rangle$ through a unitary transformation $\hat{\mathbf{ U}}_s$. Specifically, for two-state system the required unitary transformations are $\hat{\mathbf{ U}}_1=\hat{I}$, $\hat{\mathbf{ U}}_2=\mathbb{H}$, 
and $\hat{\mathbf{ U}}_3=e^{i\frac{\pi}{4}\hat{\sigma}_{x}}$ \cite{Yuan-NewJPhys2016}. Here $\mathbb{H}$ is the 
Hadamard gate, and $\hat{\sigma}_{x}=\left\vert \uparrow\right\rangle\left\langle \downarrow\right\vert+\left\vert\downarrow\right\rangle\left\langle\uparrow\right\vert$. It can be easily proved that the projective measurement 
outcome $p_{sl}$ is exactly the value of the diagonal element $\left\vert l\right\rangle\left\langle l\right\vert$ of the 
transformed density operator $\hat{\rho}_s=\hat{\mathbf{ U}}_s^{\dagger}\hat{\rho}\hat{\mathbf{ U}}_s$ \cite{Yuan-NewJPhys2016}. With these projective measurement outcomes, the MUBs-QST can be realized.
In our system, the projective measurement outcomes can be directly read out from the relative heights of the resonance 
peaks, corresponding to the computational basis state $\left\vert l\right\rangle$, in the cavity output spectra. 

Now we numerically demonstrate the MUBs-QST for a TLS in detail with above method. The density operator describing an arbitrary TLS state to be determined can be represented as $\hat{\rho}=(\hat{I}+\sum_{i}r_i\hat{\sigma}_i)/2$ ($i=x, y, z$), where $r_i$ is real parameter, and $\hat\sigma_i$ is the Pauli operator. If we choose $r_x=0.6$, $r_y=0.4$, and $r_z=0.3$, the density matrix to be measured is specified as
\begin{equation}
\rho=\left(\begin{array}{cc}0.65 & 0.3-0.2i \\0.3+0.2i & 0.35\end{array}\right).
\end{equation}
To realize the measurements in the MUBs, we first need to obtain the transformed density operator  $\hat{\rho}_s=\hat{\mathbf{ U}}_s^{\dagger}\hat{\rho}\hat{\mathbf{ U}}_s$ ($s=1,2,3$). Then we use a single-photon wave packet to implement projective measurement. According to the results in the previous section, the center 
frequency of the pulse is chosen as $\omega_0=\omega_c-\delta_1$, so that the relative heights of the double peaks 
around each sideband are proportional to the populations of states $\left\vert \uparrow\right\rangle$ and $\left\vert \downarrow\right\rangle$, respectivrly.  The corresponding 
transmission spectra around $\Delta_k=-\delta_1+\omega_b$ are shown in Figs.~\ref {state-readout}(a)-(c), from which 
we can directly read out all the projective measurement outcomes for each MUB. Specifically, in Fig.~\ref {state-readout}(a), the 
heights of the transmitted peaks marked computational basis states $\left|\downarrow\right\rangle$ and $\left|\uparrow\right\rangle$ are 0.6541 and 0.3459, respectively. Hence, the projective measurement outcomes for the MUB $|\psi_{1l}\rangle$ are $(p_{1\downarrow}, p_{1\uparrow}) = (0.6541, 0.3459)$. Similarly, we can read from Figs.~\ref {state-readout}(b) and \ref{state-readout}(c) that
$(p_{2\downarrow}, p_{2\uparrow}) = (0.8028,0.1972)$ and $(p_{3\downarrow}, p_{3\uparrow}) = (0.7037,0.2963)$, respectively. Finally, inserting these projective measurement outcomes and the MUBs  into Eq.~\eqref {density-matrix-in-MUBs}, we can obtain the reconstructed state normalized as 
\begin{equation}
\tilde{\rho}=\left(\begin{array}{cc}0.6541 & 0.3028-0.2037i \\0.3028+0.2037i & 0.3459\end{array}\right).
\end{equation}
Its fidelity can be calculated using formula
$F(\rho,\tilde{\rho})=\left[\mathrm{Tr}\left(\sqrt{\sqrt{\rho}\tilde{\rho}\sqrt{\rho}}\right)\right]^2$ \cite{Uhlmann-RepMathPhys1976,Jozsa-JModOpt1994}.  We find the fidelity is over $99.99\%$.

Here we provide a method to infer the state information of a TLS from the measured spectral data of a hybrid OMS under strong coupling regime. It is worthwhile pointing out that mechanical-motional states can also be reconstructed by using the single photon scattering spectra of a standard OMS (containing a cavity and an MR only), as studied in Ref.\cite{Liao-PRA2014}. 

\section{\label{conclusion}conclusions and discussions}  

In summary, we have explored the single-photon excitation and transmission spectra of a hybrid OMS with its MR being coupled to a TLS in the strong coupling regime. In this parameter region, the polaron-assisted excitation and reemission processes become remarkable, resulting in additional peaks in the excitation and output spectra. By analyzing in detail the fine structure around different sidebands, we find that these spectrum structures can characterize different interactions, including the TLS-MR couplings and the effective TLS-cavity coupling mediated by the MR. Thus, the hybrid quantum system studied here may provide a platform for probing the quantum nature of a macroscopic mechanical element.  Another possible application of the single-photon transmission spectra is dispersive readout of the TLS as a qubit.  Specifically, we propose a feasible proposal for quantum state tomography based on measurements in the MUBs, utilizing the hybrid OMS as a physical implementation.

\begin{acknowledgments}
S. Y. Y. and W. Z. J. were supported by the National Natural Science Foundation of China (NSFC) under Grants No. 11404269, No. 11347001, No. 61871333, No. 11747311 and No. 11947404. H. Y. was supported by the NSFC under Grant No.11405171 and the Open Foundation for CAS Key Laboratory of Quantum Information under Grant No. KQI201801.
\end{acknowledgments}

\appendix
    
\section{\label{Uele-Cal}Calculation of the matrix elements of $\hat{U}(1)$}
In this Appendix, we give the explicit formulas for the overlaps $\langle n\xi|\widetilde{n'\xi'}(1)\rangle=\langle n\xi|\hat{U}(1)|n'\xi'\rangle$, where the unitary transformation $\hat{U}(1)=\hat{U}_1(1)\hat{U}_2(1)$ is defined in Sec.~\ref{EL-structure}. Specifically, the matrix elements $\langle n\xi|\hat{U}(1)|n'\xi'\rangle$ can be written as
\begin{widetext}
\begin{subequations}
    \begin{equation}
        \left<0\downarrow\right|\hat{U}(1)\left|0\downarrow\right>=\cos{\alpha}\left<0\left|\hat{U}
        _1(1)\right|0\right>_b, 
        \label{U1}
    \end{equation}
    \begin{equation}
        \left<0\downarrow\right|\hat{U}(1)\left|n+\right>=
        \sin{\alpha}\cos{\theta_n}\left<0\left|\hat{U}_1(1)\right|n-1\right>_b
        +\cos{\alpha}\sin{\theta_n}\left<0\left|\hat{U}_1(1)\right|n\right>_b,
        \label{U2}
    \end{equation}
    \begin{equation}
        \left<0\downarrow\right|\hat{U}(1)\left|n-\right>=
        -\sin{\alpha}\sin{\theta_n}\left<0\left|\hat{U}_1(1)\right|n-1\right>_b
        +\cos{\alpha}\cos{\theta_n}\left<0\left|\hat{U}_1(1)\right|n\right>_b,
        \label{U3}
    \end{equation}
    \begin{equation}
        \left<n+\right|\hat{U}(1)\left|0\downarrow\right>=
        -\sin{\alpha}\cos{\theta_n}\left<n-1\left|\hat{U}_1(1)\right|0\right>_b
        +\cos{\alpha}\sin{\theta_n}\left<n\left|\hat{U}_1(1)\right|0\right>_b,
        \label{U4}
    \end{equation}
    \begin{equation}
        \left<n-\right|\hat{U}(1)\left|0\downarrow\right>=
        \sin{\alpha}\sin{\theta_n}\left<n-1\left|\hat{U}_1(1)\right|0\right>_b
        +\cos{\alpha}\cos{\theta_n}\left<n\left|\hat{U}_1(1)\right|0\right>_b,
        \label{U5}
    \end{equation}
    \begin{eqnarray}
        \left<n+\right|\hat{U}(1)\left|n'+\right>&=&
        \cos{\alpha}\cos{\theta_n}\cos{\theta_{n'}}\left<n-1\left|\hat{U}_1(1)\right|n'-1\right>_b
        -\sin{\alpha}\cos{\theta_n}\sin{\theta_{n'}}\left<n-1\left|\hat{U}_1(1)\right|n'\right>_b
        \nonumber 
        \\
        &&+\sin{\alpha}\sin{\theta_n}\cos{\theta_{n'}}\left<n\left|\hat{U}_1(1)\right|n'-1\right>_b
        +\cos{\alpha}\sin{\theta_n}\sin{\theta_{n'}}\left<n\left|\hat{U}_1(1)\right|n'\right>_b,
        \label{U6}
    \end{eqnarray}
    \begin{eqnarray}
        \left<n+\right|\hat{U}(1)\left|n'-\right>&=&
        -\cos{\alpha}\cos{\theta_n}\sin{\theta_{n'}}\left<n-1\left|\hat{U}_1(1)\right|n'-1\right>_b
        -\sin{\alpha}\cos{\theta_n}\cos{\theta_{n'}}\left<n-1\left|\hat{U}_1(1)\right|n'\right>_b
        \nonumber
         \\
        &&-\sin{\alpha}\sin{\theta_n}\sin{\theta_{n'}}\left<n\left|\hat{U}_1(1)\right|n'-1\right>_b
        +\cos{\alpha}\sin{\theta_n}\cos{\theta_{n'}}\left<n\left|\hat{U}_1(1)\right|n'\right>_b,
        \label{U7}
   \end{eqnarray}
   \begin{eqnarray}
        \left<n-\right|\hat{U}(1)\left|n'+\right>&=&
        -\cos{\alpha}\sin{\theta_n}\cos{\theta_{n'}}\left<n-1\left|\hat{U}_1(1)\right|n'-1\right>_b
        +\sin{\alpha}\sin{\theta_n}\sin{\theta_{n'}}\left<n-1\left|\hat{U}_1(1)\right|n'\right>_b
        \nonumber 
        \\
        &&+\sin{\alpha}\cos{\theta_n}\cos{\theta_{n'}}\left<n\left|\hat{U}_1(1)\right|n'-1\right>_b
        +\cos{\alpha}\cos{\theta_n}\sin{\theta_{n'}}\left<n\left|\hat{U}_1(1)\right|n'\right>_b,
        \label{U8}
    \end{eqnarray}
    \begin{eqnarray}
        \left<n-\right|\hat{U}(1)\left|n'-\right>&=&
        \cos{\alpha}\sin{\theta_n}\sin{\theta_{n'}}\left<n-1\left|\hat{U}_1(1)\right|n'-1\right>_b
        +\sin{\alpha}\sin{\theta_n}\cos{\theta_{n'}}\left<n-1\left|\hat{U}_1(1)\right|n'\right>_b
        \nonumber 
        \\
        &&-\sin{\alpha}\cos{\theta_n}\sin{\theta_{n'}}\left<n\left|\hat{U}_1(1)\right|n'-1\right>_b
        +\cos{\alpha}\cos{\theta_n}\cos{\theta_{n'}}\left<n\left|\hat{U}_1(1)\right|n'\right>_b,
        \label{U9}
   \end{eqnarray}
\end{subequations}
\end{widetext}
where the matrix elements of the single-photon displacement operator $\hat{U}_1(1)$ can be written as
\begin{equation}
        \left<n\left|\hat{U}_1(1)\right|n'\right>_b=
        \begin{cases}
            \sqrt{\frac{n!}{n'!}}e^{-\frac{\beta^2}{2}}(-\beta)^{n'-n}L_n^{n'-n}(\beta^2) & n'\ge n, 
            \\
            \sqrt{\frac{n'!}{n!}}e^{-\frac{\beta^2}{2}}\beta^{n-n'}L_{n'}^{n-n'}(\beta^2) & n'<n.
        \end{cases}
    \end{equation}
$L_r^s$ is the associated Laguerre polynomial. Because the matrix element $\langle n|\hat{U}_1(1)|n'\rangle_b$ is real, we can easily see from Eqs. \eqref{U1}-\eqref{U9} that the matrix elements $\langle n\xi|\widetilde{n'\xi'}(1)\rangle=\langle n\xi|\hat{U}(1)|n'\xi'\rangle$ is also real, satisfying $\langle n\xi|\widetilde{n'\xi'}(1)\rangle=\langle \widetilde{n'\xi'}(1)|n\xi\rangle$.

\section{\label{Cal-OutputSpectrum}Calculation of the transmission spectra}

We first construct the scattering matrix based on the scattering eigenstate given in Sec.~\ref {scattering-eigenstate}. According to the Lippmann-Schwinger formalism \cite{Shen-PRA2009,Sakurai-ModQuanMechanics}, Eqs. \eqref{Int-eigenstate} and \eqref {WaveFunction} tell us an input state
\begin{equation}
    \left|k,n_0\xi_0\right>=\frac{1}{\sqrt{2\pi}}\int\mathrm{d}x e^{i k x}a^{\dagger}(x)\left|\emptyset\right>\left|0\right>_c\left|n_0\xi_0\right>
\end{equation}
can be scattered to an output state
\begin{eqnarray}
    &&\left|k-{(\epsilon_{0,n\xi}-\epsilon_{0,n_{0}\xi_{0}})}/{v_{\mathrm{g}}},n\xi\right>
    \nonumber\\
    &&=\frac{1}{\sqrt{2\pi}}\int\mathrm{d}x t_{n\xi,n_0\xi_0}
    e^{i\left(k-\frac{\epsilon_{0,n\xi}-\epsilon_{0,n_{0}\xi_{0}}}{v_{\mathrm{g}}}\right)x}
    a^{\dagger}(x)\left|\emptyset\right>\left|0\right>_c\left|n\xi\right>,
    \nonumber\\
\end{eqnarray}
Here $ \left|k,n_0\xi_0\right>$ represents that initially a monochromatic single photon with wave vector $k$ incident from left, and the TLS-MR subsystem is in the state $\left|n_0\xi_0\right>$. $\left|k-{(\epsilon_{0,n\xi}-\epsilon_{0,n_{0}\xi_{0}})}/{v_{\mathrm{g}}},n\xi\right>$ is
the corresponding output state, showing that after scattering the state of the TLS-MR subsystem becomes  $\left|n\xi\right>$, and the wave vector of the scattered photon becomes $k-{(\epsilon_{0,n\xi}-\epsilon_{0,n_{0}\xi_{0}})}/{v_{\mathrm{g}}}$. Based on these results, the corresponding scattering matrix can be constructed as 
\begin{widetext}
\begin{equation}
\hat{\textbf{S}}=\sum_{n\xi}\sum_{n'\xi'}\int\mathrm{d}k t_{n\xi,n'\xi'}\left(\omega_{k}\right)|k-{(\epsilon_{0,n\xi}-\epsilon_{0,n'\xi'})}/{v_{\mathrm{g}}},n\xi\rangle \langle k,n'\xi'|.
\end{equation}
Utilizing the scattering matrix given above we can further deal with the problem of scattering of a single-photon pulse with finite bandwidth. A general incoming state can be written as
\begin{equation}
    \left|\psi_{\mathrm{in}}\right>=\sum_{n_0\xi_0}\int\mathrm{d}k f(\omega_k)C_{n_0\xi_0}\left|k,n_0\xi_0\right>,
\end{equation}
which means a single photon with spectrum amplitude $f(\omega_k)$ incident from left, and the TLS-MR subsystem is in a superposition state $\sum_{n_0\xi_0}C_{n_0\xi_0}\left|n_0\xi_0\right>$. The corresponding output state is
\begin{equation}
     \left|\psi_{\mathrm{out}}\right>=\hat{\textbf{S}}\left|\psi_{\mathrm{in}}\right>=\sum_{n\xi}\sum_{n_0\xi_0}\int\mathrm{d}k f(\omega_k)t_{n\xi,n_0\xi_0}\left(\omega_{k}\right) C_{n_0\xi_0}|k-{(\epsilon_{0,n\xi}-\epsilon_{0,n_{0}\xi_{0}})}/
     {v_{\mathrm{g}}},n\xi\rangle.
\end{equation}
After making a change of variables $\omega_k-(\epsilon_{0,n\xi}-\epsilon_{0,n_{0}\xi_{0}})\rightarrow\omega_{k}$, $k-{(\epsilon_{0,n\xi}-\epsilon_{0,n_{0}\xi_{0}})}/{v_{\mathrm{g}}}\rightarrow k$, we can find from the output sate that
\begin{equation}
    \Biggl|\sum_{n_0\xi_0} C_{n_0\xi_0}f\left(\omega_k+\epsilon_{0,n\xi}-\epsilon_{0,n_{0}\xi_{0}}\right)t_{n\xi,n_0\xi_0}\left(\omega_{k}+\epsilon_{0,n\xi}-\epsilon_{0,n_{0}\xi_{0}}
    \right)\Biggr|^2
\end{equation}
\end{widetext}
is the probability density for finding the system finally in the state  $\left|k,n\xi\right>$. Straightforwardly, to describe the probability density for finding the scattered single photon with frequency $\omega_{k}$, we can further define the transmission  spectrum $S(\omega_k)$, as shown by Eq. \eqref {ScatteringSpectrum} in the main text. 

In Sec.~\ref {state-tomography}, The transmission spectrum with the TLS-MR subsystem being prepared in the state $\hat{\rho}\otimes\left|0\right\rangle_b\left\langle0\right|_b$ was used for quantum state tomography. Here $\hat{\rho}=\sum_{ij=\downarrow,\uparrow}\rho_{ij}\left|i\right\rangle\left\langle j\right|$ is the density operator (usually representing a mixed state) of the TLS (qubit). Utilizing an appropriate unitary transformation, the density operator can be diagonalized: $\hat{\rho}=\sum_{u}\mathcal{P}_{u}\left|\phi_{u}\right\rangle\left\langle \phi_{u}\right|$, with $\mathcal{P}_{u}=\sum_{ij}\mathcal{(U^{\dagger})}_{ui}\rho_{ij}\mathcal{U}_{ju}$ and $\left|\phi_{u}\right\rangle=\sum_{i}\mathcal{U}_{iu}\left|i\right\rangle$. Clearly, the transmission spectrum corresponding to the initial state $\hat{\rho}\otimes\left|0\right\rangle_b\left\langle0\right|_b$ can be written as  $S\left(\omega_{k}\right)=\sum_u\mathcal{P}_{u}S_u\left(\omega_{k}\right)$. Here, $S_u\left(\omega_{k}\right)$ is the output spectrum with the subsystem in the pure state $\left|0\right\rangle_b\left|\phi_{u}\right\rangle$, and can be calculated by Eq.~\eqref{ScatteringSpectrum}.

\bibliography{MS-Aug-15-2020}

\end{document}